\documentclass[conference, 10pt]{IEEEtran}
\usepackage{amssymb}
\usepackage{amsmath, tabularx}
\usepackage{mathtools}
\usepackage{ellipsis}
\usepackage{textcomp}
\usepackage{breqn}
\usepackage{cite}
\usepackage{graphicx}	
\usepackage{subfigure}
\usepackage{epsfig}
\usepackage{caption}
\usepackage{psfrag}
\usepackage{grffile}

\usepackage{setspace}
\usepackage{epstopdf} 
\usepackage{color}
\usepackage{multirow}
\usepackage{amsthm}
\newtheorem{mylemma}{Lemma}

\usepackage[overload]{empheq}
\usepackage{cases}
\usepackage{url}

\usepackage{pifont}
\newcommand{\cmark}{\ding{51}}%
\newcommand{\xmark}{\ding{55}}%

\usepackage{xcolor}
\usepackage[noend]{algpseudocode}
\usepackage{algorithm}
\algnewcommand{\LineComment}[1]{\Statex \(\triangleright\) #1} 
\algnewcommand{\And}{\textbf{and}}
\algnewcommand{\Or}{\textbf{or}}
\usepackage{multicol}

\algnewcommand{\LineCommentCont}[1]{\Statex \hskip\ALG@thistlm%
  \parbox[t]{\dimexpr\linewidth-\ALG@thistlm}{\hangindent=\trianglerightwidth \hangafter=1 \strut$\triangleright$ #1\strut}}

\DeclareTextFontCommand{\textOF}{\fontfamily{lmtt}\selectfont}

\IEEEoverridecommandlockouts
\begin{document}




\title{Availability-Aware VNF Placement and Request Routing in MEC-Enabled 5G Networks}
\author{Work in Progress}
\author{Aqsa~Sayeed and Samaresh~Bera,~\textit{Senior~Member,~IEEE}
\thanks{The preliminary version of this work has been accepted for publication in IEEE ANTS 2023~\cite{beraAvailabilityawareVNFPlacement2023}.}
\thanks{The authors are with the Department of Computer Science and Engineering, Indian Institute of Technology Jammu, 181221, India. Email:~s.bera.1989@ieee.org.}
}

\maketitle

\begin{abstract}
In this paper, we study the virtual network function (VNF) placement problem in mobile edge computing (MEC)-enabled 5G networks to meet the stringent reliability and latency requirements of uRLLC applications. We pose it as a constrained optimization problem, which is NP-hard, to maximize the total reward obtained by a network service provider by serving uRLLC service requests. We propose an approximated randomized rounding approach to solve the NP-hard optimization problem in polynomial time. We prove that the proposed randomized approach achieves performance guarantees while violating the resource constraints boundedly. Furthermore, we present a greedy-heuristic approach to tackle the violations of resource constraints. 

Simulation results show that the proposed randomized rounding and greedy approaches achieve a total reward which is within 5\% and 10\% of the optimal solution, respectively. Furthermore, we compare the proposed greedy approach with the existing schemes that do not consider the availability requirements. We observe that the existing schemes perform poorly in terms of total reward, as negligence to the availability requirements negatively impacts the number of successfully served requests. These findings highlight the trade-off between availability and resource efficiency in latency-sensitive uRLLC applications. We also implement a software prototype of a 5G network using open-source software platforms with redundant placement of VNFs. The results on packet delivery ratio and latency obtained from the prototype implementation are also improved in the redundant VNFs with different failure probabilities.

\end{abstract}


\section{Introduction} \label{Sec:Introduction}


The emergence of the mobile edge computing (MEC) framework facilitates network service providers to meet the stringent latency requirements of real-time applications by placing computing functionalities near the users. Furthermore, the concept of MEC has proliferated with the introduction of 5G (and beyond) networks and its use-cases, such as autonomous driving and remote surgery. With MEC, a service can be hosted either at the edge cloud or at the central cloud, depending on its requirements. Recent studies show that MEC is helpful in meeting the stringent latency requirements of the above-mentioned applications while leveraging the benefits of software-defined networking (SDN) and network function virtualization (NFV)~\cite{yalaLatencyAvailabilityDriven2018, sarrigiannisOnlineVNFLifecycle2020, harutyunyanCostefficientPlacementScaling2021, poularakisJointServicePlacement2019}. The SDN and NFV technologies help the network operator to place service-specific softwarized network functions, called virtual network functions (VNFs), in the network to reduce the CAPEX and OPEX~\cite{yousafNFVSDNKey2017a}.

The applications supported by 5G (and beyond) are broadly categorized as enhanced mobile broadband (eMBB), ultra-reliable and low-latency communications (uRLLC), and massive machine-type communications (mMTC)~\cite{5GProgrammableInfrastructure2018, huangMECEnabledTaskReplication2025, siewEffectiveResourceProcurement2024}. The eMBB, uRLLC, and mMTC applications require network support in terms of high bandwidth, high reliability and low-latency, and massive number of devices, respectively. While there has been significant progress in addressing the high-bandwidth and low-latency requirements of 5G applications~\cite{NavarroOrtizSurvey5GUsage2020}, ensuring high reliability is still challenging for uRLLC applications. This is due to the coupling of high-reliability and low-latency requirements of uRLLC applications that makes the network modeling very challenging~\cite{popovskiWirelessAccessUltrareliable2019}. Considering this, in this work, we ask the following questions:

\begin{itemize}
	\item Which uRLLC service requests should be served by MEC instead of the central cloud considering networking resources and request-specific requirements?
	\item How to route user traffic of the service requests to the MEC, if selected, while considering uplink and downlink networking resources to the MECs?
	\item How to optimize the above decisions to maximize total reward by serving requests at the MECs, so that the underlying applications experience a lower service latency?
\end{itemize}

To answer the above-mentioned questions, in this paper, we study the VNF placement and request routing problem in an MEC-enabled 5G network. We pose this as a constrained optimization problem, which is NP hard. The optimization problem captures multi-dimensional networking resources, such as CPU, RAM, uplink, and downlink resources, to maximize the total reward obtained by the service provider. In addition to the multi-dimensional networking resources, the problem considers the availability requirements of the service requests while placing the VNFs associated with the requests. Furthermore, we propose an approximation algorithm to solve the NP-hard optimization problem with performance bounds. The key contributions in this paper are as follows:
\begin{itemize}
	\item We mathematically model the VNF placement and request routing as a constrained optimization problem to maximize the total reward to a service provider by serving requests. The problem considers multi-dimensional networking resources and availability requirements to meet the stringent reliability of 5G applications.
	
	\item We propose an approximation algorithm based on randomized rounding techniques~\cite{motwaniRandomizedAlgorithms1995} to solve the NP-hard optimization problem in polynomial time. Furthermore, we show that it provably achieves performance guarantees while violating the resource constraints in a bounded way.
	
	\item We propose a greedy-heuristic approach based on the randomized rounding solution to tackle the violations in resource constraints, if any. The extensive results show that the proposed approximation algorithm performs close-to-optimal while violating the constraints in a bounded way. Furthermore, the greedy approach yields competitive performance to the optimal solution. The proposed greedy approach also outperforms the existing scheme that do not consider the availability requirement. 
	
	\item We implement a software prototype of the 5G network and place redundant VNFs using open-source software platforms. We present the network performances on packet delivery ratio and latency in the presence of redundant VNFs with different failure probabilities.
\end{itemize}

The rest of the paper is organized as follows. Section~\ref{Sec:Related_work} highlights the state-of-the-art solution approaches for the VNF placement and request routing problem. Section~\ref{Sec:System_model} presents the detailed network model and the optimization problem. Section~\ref{Sec:Randomized_rounding} presents the proposed approximation algorithm while analyzing the bounds on the performance, and the proposed greedy approach. The efficacy of the proposed scheme is studied in Section~\ref{Sec:Performance_evaluation}. Finally, Section~\ref{Sec:Conclusion} concludes the paper with future research directions.

\section{Related Work}\label{Sec:Related_work}
This section discusses the existing works on VNF placement and resource allocation while highlighting the key differences between them and the proposed scheme. We categorize the existing works on VNF placement as follows: a) non-5G network and b) 5G network. The works on the 5G network are further divided into two categories based on the network functionalities -- i)~5G radio access network (RAN) and ii)~5G core network.



\subsection{VNF Placement in Non-5G Network}\label{Secsub:related_non_5g_network}
Poularakis et al.~\cite{poularakisJointServicePlacement2019, poularakisServicePlacementRequest2020} studied service placement and request routing problem in an MEC-enabled network, where base-stations are enabled with storage and compute resources and act as edge clouds. The authors framed the optimization problem as the minimization of request routing to the central cloud while adhering to the associated constraints. Similarly, Yang et al.~\cite{yangDelayawareVirtualNetwork2021} framed the VNF placement and routing problem as the minimization of service delay while considering the networking resources and request-specific requirements. Both the works~\cite{poularakisJointServicePlacement2019} and \cite{yangDelayawareVirtualNetwork2021} proposed approximation algorithms to solve the NP-hard optimization problems in polynomial time. 

\subsection{VNF Placement in 5G Network}\label{Secsub:related_5g_network}

\subsubsection{5G Radio Access Network (RAN)}\label{Secsubsub:related_5g_RAN}

Marotta et al.~\cite{marottaImpactCoMPVNF2017} studied the impact of VNF placement at the 5G RAN where multiple base stations coordinate to increase throughput. The authors emulated different coordinated multipoint (CoMP) scenarios of base stations at the RAN and evaluated their impact on throughput. They suggested \textit{optimal} deployment scenarios based on the findings. In such a scenario, the VNFs are used to schedule the physical resource blocks allocation to a user from multiple base stations so that interference from cooperating base stations can be reduced.

Behravesh et al.~\cite{behraveshJointUserAssociation2019} studied a joint user association and VNF placement problem in the network consisting of MECs and the central cloud. The authors formulated it as a mixed integer linear program (MILP) and solved it using an optimization problem solver. However, it is unsuitable for large-scale deployment due to the NP-hardness of the optimization problem.

\begin{table*}[!ht]
	\centering
	\caption{Key differences between the proposed scheme and existing schemes}
	\label{Table:related_work}
	\renewcommand{\arraystretch}{1.2}
	\begin{tabular}{lccccccccc}
		\hline
		\multirow{2}{*}{\textbf{Works}} & \multirow{2}{*}{\textbf{MEC}} & \multicolumn{4}{c}{\textbf{Networking resources}}                 & \multirow{2}{*}{\textbf{\begin{tabular}[c]{@{}c@{}}VNF\\ Placement\end{tabular}}} & \multirow{2}{*}{\textbf{Avail.}} & \multirow{2}{*}{\textbf{\begin{tabular}[c]{@{}c@{}}Redundancy\\ type\end{tabular}}} \\ \cline{3-6}
		&                                     & \textbf{CPU} & \textbf{RAM} & \textbf{Uplink} & \textbf{Downlink} &                                                                                   &                                   &                                                                                                                          \\ \hline
		Behravesh et al.~\cite{behraveshJointUserAssociation2019}                       & \cmark                                   & \cmark            & \xmark            & \xmark               & \xmark                 & \cmark                                                                                                                & \cmark                                      & Not considered                                                                      \\ \hline
		Yala et al.~\cite{yalaLatencyAvailabilityDriven2018}                           & \cmark                                   & \cmark            & \xmark            & \xmark               & \xmark                 & \cmark                                                                                                                 & \cmark                                      & active-passive                                                                      \\ \hline
		Poularakis et al.~\cite{poularakisJointServicePlacement2019, poularakisServicePlacementRequest2020}                     & \cmark                                   & \cmark           & \cmark            & \cmark              & \cmark                 & \cmark                                                                                                                & \xmark                                      & Not considered                                                                      \\ \hline
		Yang et al.~\cite{yangDelayawareVirtualNetwork2021}                           & \cmark                                   & \cmark            & \xmark            & \cmark               & \cmark                 & \cmark                                                                                                                & \xmark                                      & Not considered                                                                      \\ \hline
		\textbf{Proposed}                        & \cmark                                   & \cmark            & \cmark            & \cmark               & \cmark                 & \cmark                                                                                                                  & \cmark                                      & \textbf{active-active}                                                                       \\ \hline
	\end{tabular}
\end{table*}

\subsubsection{5G Core Network}\label{Secsubsub:related_5g_core}

Yala et al.~\cite{yalaLatencyAvailabilityDriven2018} studied availability and latency-aware VNF placement problem at MECs and the central cloud. The authors modeled it as a trade-off between the service latency and the availability of VNFs. For latency-critical services, the placement of VNFs is preferred at the MECs over the central cloud. In contrast, the VNFs related to availability-critical services are placed in the central cloud. However, as discussed in~\cite{VerticalsURLLCUse2020}, the uRLLC use cases in 5G have stringent latency and availability requirements, which must be ensured. Furthermore, the authors propose a genetic algorithm-based approach for which the solution may not converge even after multiple iterations.


Network slicing is a viable approach for meeting diverse performance requirements of different 5G use-cases~\cite{rostNetworkSlicingEnable2017, afolabiNetworkSlicingSoftwarization2018} by creating multiple logical networks. A VNF can be shared among multiple slices to reduce deployment and operational costs. However, in such a scenario, performance may be degraded for specific service(s) due to the traffic from other service requests, called as \textit{interference}. In this context, Zhang et al.~\cite{zhangAdaptiveInterferenceawareVNF2019} studied the interference-aware VNF placement problem to maximize total throughput. Mohan and Gurusamy ~\cite{mohanResilientVNFPlacement2019a} studied the VNF placement problem for network slices with diverse performance requirements. The authors proposed a resilient VNF placement approach while considering two types of VNF failures -- fine-grained and course-grained. In fine-grained, a VNF fails due to the software failure in the VNF itself. In contrast, course-grained failures happen due to the failure of physical machines.

\textit{Synthesis}: While~\cite{poularakisJointServicePlacement2019, poularakisServicePlacementRequest2020, yangDelayawareVirtualNetwork2021} are the closest ones to our work, they did not consider the availability of VNFs, which makes the problem challenging due to the limited networking resources at the MECs. Furthermore, while the existing schemes tried to address the issues and challenges in supporting the stringent QoS requirements, very few are scalable and yield competitive performance to the optimal solution. In summary, Table~\ref{Table:related_work} presents the key differences between the proposed and existing schemes. It is evident that research lacuna exists in VNF placement and request routing at the MECs considering the high-reliability requirements of 5G applications.

\section{System Model}\label{Sec:System_model}
\begin{figure}[!ht]
	\centering
	\includegraphics[scale=1.35]{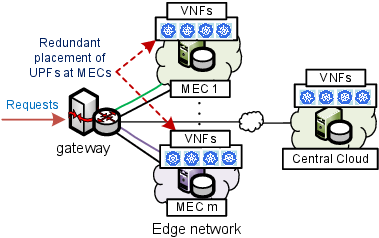}
	\caption{Network model: MEC-based VNF placement}
	\label{fig:Architecture}
\end{figure}

Figure~\ref{fig:Architecture} presents an overview of the MEC-assisted VNF placement in the network. The VNFs associated with a request are placed at the MECs or the central cloud. Furthermore, the number of redundant VNF placements is calculated based on the availability requirement by the request. Table~\ref{Table:list_of_symbols} presents the list of symbols used in this work.

\begin{table}[!ht]
	\centering
	\caption{List of symbols}
	\label{Table:list_of_symbols}
	\renewcommand{\arraystretch}{1.25}
	\begin{tabular}{l p{7cm}}
		\textbf{Symbol} & \textbf{Detail}\\ \hline
		$\mathcal{M}$ & Set of MECs in the network\\ \hline
		$M$ & Number of MECs, i.e., $M = |\mathcal{M}|$\\ \hline
		$\mathcal{R}$ & Set of service requests\\ \hline
		$R$ & Number of requests, i.e., $R = |\mathcal{R}|$\\ \hline
		$C_m$ & CPU capacity of an MEC $m \in \mathcal{M}$\\ \hline
		$D_m$ & RAM capacity of an MEC $m \in \mathcal{M}$\\ \hline
		$B_{m}^{\text{up}}$ & Uplink capacity for routing to MEC $m \in  \mathcal{M}$\\ \hline
		$B_{m}^{\text{dw}}$ & Uplink capacity for routing from MEC $m \in  \mathcal{M}$\\ \hline
		$\epsilon_{\text{m}}$ & Service failure probability from an MEC $m \in \mathcal{M}$\\ \hline
		$\epsilon_{\text{p}}$ & Physical failure probability of a UPF\\ \hline
		$\epsilon_{\text{v}}$ & Software failure probability of a UPF\\ \hline
		$c_r$ & CPU requirement of request $r \in \mathcal{R}$\\ \hline
		$d_r$ & RAM requirement of request $r \in \mathcal{R}$\\ \hline
		$b_{r}^{\text{up}}$ & Uplink requirement of request $r \in \mathcal{R}$\\ \hline 
		$b_{r}^{\text{dw}}$ & Downlink requirement of request $r \in \mathcal{R}$\\ \hline 
		$\epsilon_r$ & Failure threshold requirement of request $r \in \mathcal{R}$\\ \hline
		$\zeta_r$ & Reward by serving request $r \in \mathcal{R}$\\ \hline
		$x_{r, m}$ & Denotes whether VNFs associated with request $r \in \mathcal{R}$ are placed at MEC $m \in \mathcal{M}$\\ \hline
		$y_r$ & Denotes whether request $r \in \mathcal{M}$ is served\\ \hline
	\end{tabular}
\end{table}

In this work, we consider the following:
\begin{itemize}
	
	\item VNFs in 5G network are categorized as control-plane functions (CPFs) and user-plane functions (UPFs). The CPFs are pre-placed either at the MEC, central cloud, or both.
	
	\item The VNFs are placed in the form of a virtual machine (VM) or a container. Furthermore, we focus on the UPF\footnote{In this work, UPFs and VNFs are used interchangeably henceforth, unless specified otherwise, as our primary focus is on the VNF placement at the user-plane of a 5G network.} placement at the MECs in the network. The requests that cannot be served by MEC are either dropped or served by central cloud while considering their requirements.
	
	\item Considering issues with latency in serialized HTTP/JSON
	connection~\cite{jainL25GCLowLatency2022}, all UPFs associated with a service are placed at a single physical machine (PM).
	
	\item A request is replicated and assigned to multiple
	MECs for providing services based on the availability requirement.
	
	\item The first error-free service response returns to the gateway is considered. Hence, end-to-end latency and availability are improved.
\end{itemize}

\subsection{Network Model}\label{Secsub:System_network_model}
Let there be a set of MECs denoted as $\mathcal{M} = \{1, 2, \cdots, M\}$. Each MEC $m \in \mathcal{M}$ has a certain amount of CPU and RAM resources to host UPFs, denoted by $C_m$ and $D_m$, respectively. Furthermore, each MEC has an uplink and downlink capacity that can be used to route data from/to gateway to/from it. The uplink and downlink capacities are denoted by $B_m^{\text{up}}$ and $B_m^{\text{dw}}$, $\forall m \in \mathcal{M}$, respectively. The service provider receives service requests denoted by a set $\mathcal{R}= \{1, 2, \cdots, R\}$. Each request $r \in \mathcal{R}$ has a certain amount of CPU and RAM resource requirements according to the UPFs associated with the service, denoted by $c_r$ and $d_r$, respectively. Furthermore, each request is associated with a threshold on service failure $\epsilon_r$ to meet availability requirements, uplink $b_{r}^{\text{up}}$ and downlink $b_{r}^{\text{dw}}$ capacity requirements, and a reward $\zeta_r$ obtained by the service provider if the request is served.

\subsection{Latency Model}\label{Secsub:System_latency_model}
As shown in Figure~\ref{fig:Architecture}, a request can be served either at the MECs or at the central cloud. We also consider that the service execution rate at a UPF is independent of its placement (whether at MEC or the central cloud) when the same amount of resources are allocated to the UPF. Therefore, in this work, we consider that the placement of UPFs at the MECs helps in reducing the end-to-end latency similar to the existing works~\cite{yalaLatencyAvailabilityDriven2018, yangDelayawareVirtualNetwork2021, harrisDynamicVNFPlacement2022}. For brevity, we limit our discussion on the computation delay at the VNFs. Interested readers may refer to~\cite{chenTaskOffloadingMobile2018, misraSoftVANMobilityawareTask2020} for details.

Considering the low-latency requirements of uRLLC applications, we aim to serve the requests by placing UPFs at the MECs as much as possible to reduce the service delay.

\subsection{Availability Model}\label{Secsub:System_availability_model}
A service may not be available due to the following reasons:
\begin{itemize}
	\item At least one UPF associated with the service fails due to software failure of the UPF itself. Let the probability of a UPF failure be $\epsilon_{\text{v}}$.
	\item A UPF fails due to the failure of the host physical machine. Let the probability of a physical machine failure be $\epsilon_{\text{p}}$.
\end{itemize}

As all the UPFs associated with a request are placed at the same MEC, the total probability of failure of a service is calculated as:
\begin{align*}
	\epsilon_{\text{m}} & = \text{Pr}\left[ \text{UPF fails} | \text{PM is OK} \right] + \text{Pr}\left[ \text{UPF fails} | \text{PM fails} \right]\notag \\
	& =  \left(\epsilon_{\text{v}} + \epsilon_{\text{p}} \right).
\end{align*}

Now, for a given request $r \in \mathcal{R}$ with a threshold failure requirement $\epsilon_r$, the following condition needs to be satisfied to meet the availability requirement:
\begin{equation}
	\left[ \epsilon_{\text{m}} \right]^{\Psi_r} \leq \epsilon_r,
	\label{eqn:availability_requirement_pre}
\end{equation}
where $\Psi_r$ denotes the number of redundant placement of UPFs at MECs for request $r$. To get the number of redundant placement $\Psi_r$, we rewrite~\eqref{eqn:availability_requirement_pre} as follows:
\begin{equation}
	\Psi_r = \lceil\log_{\epsilon_{\text{m}}}{(\epsilon_r)}\rceil.
	\label{eqn:availability_requirement}
\end{equation}

\subsection{Problem Statement}\label{Secsub:System_problem_statement}
Given the MECs with resources and the requests with requirements, the objective of the service provider is to maximize the total reward by serving the requests at the MECs while adhering to the associated constraints. Mathematically,
\begin{equation}
	\text{Maximize }	P_{\text{IP}} = \sum_{r \in \mathcal{R}} \zeta_r y_r,
	\label{eqn:objective_optimization}
\end{equation}

subject to

\begin{subequations}
	\begin{alignat}{2}
		& \sum_{m \in \mathcal{M}} x_{r, m} \geq \Psi_r y_r, \forall r\in \mathcal{R}, \label{eqn:constraint_redundancy}\\
		& y_r \leq 1, \forall r \in \mathcal{R},	\label{eqn:constraint_max_service}\\
		& \sum_{r \in \mathcal{R}} c_r x_{r, m} \leq C_m, \forall m \in \mathcal{M}, \label{eqn:constraint_cpu_capacity}\\
		& \sum_{r \in \mathcal{R}} d_r x_{r, m} \leq D_m, \forall m \in \mathcal{M}, \label{eqn:constraint_ram_capacity}\\
		& \sum_{r \in \mathcal{R}} b_{r}^{\text{up}} x_{r, m} \leq B_m^{\text{up}}, \forall m \in \mathcal{M}, \label{eqn:constraint_uplink_capacity}\\
		& \sum_{r \in \mathcal{R}} b_{r}^{\text{dw}} x_{r, m} \leq B_m^{\text{dw}}, \forall m \in \mathcal{M}, \label{eqn:constraint_downlink_capacity}\\
		& x_{r, m} \text{ and } y_r \in \{0, 1\}, \forall r \in \mathcal{R}, \forall m \in \mathcal{M}. \label{eqn:constraint_binary_variable}
	\end{alignat}
\end{subequations}

Equation~\eqref{eqn:objective_optimization} denotes the objective to maximize the total reward by serving requests at the MECs, so that service latency is also minimized. The service requests that cannot be served at the MEC are either dropped or forwarded to the central cloud. Equation~\eqref{eqn:constraint_redundancy} ensures that the availability requirement is satisfied. Furthermore, it also ensures that if a request is served, UPFs associated with it must be placed in the network. A request can be admitted at most once, which is ensured in~\eqref{eqn:constraint_max_service}. Equations~\eqref{eqn:constraint_cpu_capacity} and \eqref{eqn:constraint_ram_capacity} present that the CPU and RAM utilizations at the MECs are within the total CPU and RAM capacities, respectively. Equations~\eqref{eqn:constraint_uplink_capacity} and \eqref{eqn:constraint_downlink_capacity} present the uplink and downlink capacity constraints, respectively. Finally, Equation~\eqref{eqn:constraint_binary_variable} represents the binary decision variables, where $x_{r, m} = 1$ if UPFs associated with request $r$ are placed at MEC $m$, else $0$; and $y_{r} = 1$ is the request $r$ is served, else $0$. The optimization problem is a variation of the multi-constraint knapsack problem, which is NP-hard in general~\cite{martelloKnapsackProblemsAlgorithms1990}. In the subsequent section, we propose a polynomial time approximation algorithm to solve this problem in polynomial time.

\section{Solution Approach: Randomized Rounding}\label{Sec:Randomized_rounding}

\subsection{Approximated Solution}\label{Secsub:approximated_solution}
We first relax the binary variables to continuous ones to solve the problem~\eqref{eqn:objective_optimization} in polynomial time. Mathematically, the optimization problem is represented as:
\begin{equation}
	\text{Maximize }	P_{\text{LR}} = \sum_{r \in \mathcal{R}} \zeta_r y_r,
	\label{eqn:objective_optimization_relax}
\end{equation}

subject to

\begin{subequations}
	\begin{alignat}{2}
		& \eqref{eqn:constraint_redundancy}, \eqref{eqn:constraint_max_service}, \eqref{eqn:constraint_cpu_capacity},  \eqref{eqn:constraint_ram_capacity}, \eqref{eqn:constraint_uplink_capacity}, \text{ and } \eqref{eqn:constraint_downlink_capacity} \notag\\
		& x_{r, m} \text{ and } y_r \in [0, 1]. \label{eqn:constraint_continuous_variable}
	\end{alignat}
\end{subequations}

The problem in~\eqref{eqn:objective_optimization_relax} can be solved using standard LP-solvers. We use the \textOF{IBM CPLEX}~\cite{CPLEXCPOptimizer2020} to get the solution. Let the solution be $\tilde{\mathbf{x}}$ and $\tilde{\mathbf{y}}$. Now, we round the solution of the relaxed problem using the randomized rounding approach~\cite{motwaniRandomizedAlgorithms1995}, as presented in Algorithm~\ref{Algo1:Randomized_rounding}.

\begin{algorithm}[!ht]
	\begin{algorithmic}[1]
		\Require{Set of MECs: $\mathcal{M}$,\newline
			\hspace*{1.5cm} each with $C_m$, $D_m$, $B_{m}^{\text{up}}$, $B_{m}^{\text{dw}}$, $\forall m \in \mathcal{M}$;\newline
		\hspace*{0.7cm}Set of requests: $\mathcal{R}$, \newline
		\hspace*{1.5cm} each with $c_r$, $d_r$, $b_{r}^{\text{up}}$, $b_{r}^{\text{dw}}$, $\zeta_r$, and $\epsilon_r$, $\forall r \in \mathcal{R}$;
		}
		\Ensure{Binary solution: $\hat{\mathbf{x}}$ and $\hat{\mathbf{y}}$
		}
		
		\State Calculate $\Psi_r$ using~\eqref{eqn:availability_requirement}
		\State Solve the optimization problem in~\eqref{eqn:objective_optimization_relax} to obtain $(\tilde{\mathbf{x}}, \tilde{\mathbf{y}})$
		
		\For {$r \in \mathcal{R}$}
			\For {$m \in \mathcal{M}$}
				\State Set $\hat{x}_{r, m} = 1$ with probability $\tilde{x}_{r, m}$ \label{step:x_equal_one_with_probability}
				\Statex \hspace*{1.4cm} and $\hat{x}_{r, m} = 0$ with probability $(1 - \tilde{x}_{r, m})$
			\EndFor
			
			\If{$\sum\limits_{m \in \mathcal{M}} \hat{x}_{r, m} \geq \Psi_r$}\label{step:Algo_rounding_redundancy_check}
				\State Set $\hat{y}_r = 1$ with probability $\tilde{y}_r$
				\Statex \hspace*{1.5cm} and $\hat{y}_r = 0$ with probability $(1 - \tilde{y}_r)$
			\EndIf
		\EndFor
		\State \Return $(\hat{\mathbf{x}}, \hat{\mathbf{y}})$
	\end{algorithmic}
	\caption{Randomized rounding algorithm}
	\label{Algo1:Randomized_rounding}
\end{algorithm}

From the construction of Algorithm~\ref{Algo1:Randomized_rounding}, a request is either served at the MECs or dropped (or can be served by the central cloud). Therefore, the constraint \eqref{eqn:constraint_max_service} is always satisfied. Furthermore, the \textOF{Step}~\ref{step:Algo_rounding_redundancy_check} satisfies the redundancy constraint~\eqref{eqn:constraint_redundancy} to meet the availability requirements. Now, we check whether the remaining constraints~\eqref{eqn:constraint_cpu_capacity}, \eqref{eqn:constraint_ram_capacity}, \eqref{eqn:constraint_uplink_capacity}, and \eqref{eqn:constraint_downlink_capacity} are satisfied.

\begin{mylemma}\label{lemma:cpu_capacity_expectation}
	The solution returned by Algorithm~\ref{Algo1:Randomized_rounding} satisfies the CPU, RAM, uplink, and downlink capacity constraints in expectation.
\end{mylemma}

\begin{proof}
	
	\textbf{CPU capacity constraint:} The expected CPU utilization of an MEC $m \in \mathcal{M}$ is given by
	
	\begin{multline}
		\mathbb{E}\left[ \sum_{r \in \mathcal{R}} \hat{x}_{r, m} c_r \right] = \sum_{r \in \mathcal{R}} \text{Pr} \left[ \hat{x}_{r, m} = 1 \right] c_r\\ = \sum_{r \in \mathcal{R}} \tilde{x}_{r, m} c_r \leq C_m,
		\label{eqn:cpu_capacity_expectation}
	\end{multline}
	where the last equality holds as $\{\hat{x}_{r, m}\} = 1$ with success probabilities $\{\tilde{x}_{r, m}\}$ (refer to \textOF{Step}~\ref{step:x_equal_one_with_probability}). Furthermore, the inequality holds due to the constraint~\eqref{eqn:constraint_cpu_capacity}.
	
	\textbf{RAM capacity constraint:} The expected RAM utilization of an MEC $m \in \mathcal{M}$ is given by
		\begin{multline}
			\mathbb{E}\left[ \sum_{r \in \mathcal{R}} \hat{x}_{r, m} d_r \right] = \sum_{r \in \mathcal{R}} \text{Pr} \left[ \hat{x}_{r, m} = 1 \right] d_r\\ = \sum_{r \in \mathcal{R}} \tilde{x}_{r, m} d_r \leq D_m.
			\label{eqn:ram_capacity_expectation}
		\end{multline}
	Similar to~\eqref{eqn:cpu_capacity_expectation}, the last equality and the inequality hold due to the success probabilities $\{\tilde{x}_{r, m}\}$ and constraint~\eqref{eqn:constraint_ram_capacity}, respectively.

	\textbf{Uplink capacity constraint:} The expected uplink utilization to route traffic to MEC $m \in \mathcal{M}$ is given by
	\begin{multline}
		\mathbb{E}\left[ \sum_{r \in \mathcal{R}} \hat{x}_{r, m} b_{r}^{\text{up}} \right] = \sum_{r \in \mathcal{R}} \text{Pr} \left[ \hat{x}_{r, m} = 1 \right] b_{r}^{\text{up}}\\ = \sum_{r \in \mathcal{R}} \tilde{x}_{r, m} b_{r}^{\text{up}} \leq B_{m}^{\text{up}}.
		\label{eqn:uplink_capacity_expectation}
	\end{multline}
	Similar to~\eqref{eqn:cpu_capacity_expectation}, the last equality and the inequality hold due to the success probabilities $\{\tilde{x}_{r, m}\}$ and constraint~\eqref{eqn:constraint_uplink_capacity}, respectively.

	\textbf{Downlink capacity constraint:} The expected downlink utilization to route traffic from MEC $m \in \mathcal{M}$ is given by
	\begin{multline}
		\mathbb{E}\left[ \sum_{r \in \mathcal{R}} \hat{x}_{r, m} b_{r}^{\text{dw}} \right] = \sum_{r \in \mathcal{R}} \text{Pr} \left[ \hat{x}_{r, m} = 1 \right] b_{r}^{\text{dw}}\\ = \sum_{r \in \mathcal{R}} \tilde{x}_{r, m} b_{r}^{\text{dw}} \leq B_{m}^{\text{dw}}.
		\label{eqn:downlink_capacity_expectation}
	\end{multline}
	Similar to~\eqref{eqn:cpu_capacity_expectation}, the last equality and the inequality hold due to the success probabilities $\{\tilde{x}_{r, m}\}$ and constraint~\eqref{eqn:constraint_downlink_capacity}, respectively.
\end{proof}

\begin{mylemma}
	The objective value returned by Algorithm~\ref{Algo1:Randomized_rounding} is in expectation equal to that of the optimal fractional solution.
\end{mylemma}

\begin{proof}
	The expected reward obtained by the service provider by serving requests at the MECs is given by
		\begin{equation}
			\mathbb{E}\left[ \sum_{r \in \mathcal{R}} \zeta_r \hat{y}_r \right] = \sum_{r \in \mathcal{R}} \text{Pr}\left[ \hat{y}_r = 1 \right] \zeta_r = \sum_{r \in \mathcal{R}} \tilde{y}_r \zeta_r,
			\label{eqn:objective_expectation}
		\end{equation}
	where the last equality holds as $\{\hat{y}_r = 1\}$ with success probabilities $\{\tilde{y}_r\}$.
\end{proof}

Though the constraints~\eqref{eqn:constraint_cpu_capacity}, \eqref{eqn:constraint_ram_capacity}, \eqref{eqn:constraint_uplink_capacity}, and \eqref{eqn:constraint_downlink_capacity} are satisfied in expectation, they can be violated in practice. Therefore, we give the theoretical bounds on the violation of the constraints below.

\begin{mylemma}\label{lemma:CPU_capacity}
	The CPU load on an MEC $m \in \mathcal{M}$ returned by the Algorithm~\ref{Algo1:Randomized_rounding} will not exceed its capacity by more than a factor of $(1 + \delta_{\text{c}}) = \frac{3 \ln \left(R\right)}{\mu_{\text{c}}} + 4$ with high probability, where $\mu_{\text{c}} = \frac{\sum\limits_{r \in \mathcal{R}} \tilde{x}_{r, m} c_r}{\alpha_{\text{c}}}$ and $\alpha_{\text{c}} = \max\{c_r, \forall r \in \mathcal{R}\}$.
\end{mylemma}

\begin{proof}
	We are interested in finding the probability that the constraint is violated by a factor of $\delta_{\text{c}} \geq 0$. Mathematically,
	\begin{equation}
		\text{Pr}\left[ \sum_{r \in \mathcal{R}} \hat{x}_{r, m} c_r \geq \left(1 + \delta_{\text{c}} \right) \sum_{r \in \mathcal{R}} \tilde{x}_{r, m} c_r \right].
		\label{eqn:lemma_math_expression_violation_cpu}
	\end{equation}

	We will apply the Chernoff Bound~\cite{mitzenmacherProbabilityComputingRandomized2005} to get a theoretical bound on the above probability. Before that, we normalize the expression in~\eqref{eqn:lemma_math_expression_violation_cpu}, which is given by:
	\begin{align}
		& \text{Pr}\left[ \sum_{r \in \mathcal{R}} \frac{\hat{x}_{r, m} c_r}{\alpha_{\text{c}}} \geq \left(1 + \delta_{\text{c}} \right) \sum_{r \in \mathcal{R}} \frac{\tilde{x}_{r, m} c_r}{\alpha_{\text{c}}} \right],\\ 
		& \hspace*{2cm} \text{ where } \alpha_{\text{c}} = \max\{c_r, \forall r \in \mathcal{R}\},\notag \\
		& \Rightarrow \text{Pr}\left[ \sum_{r \in \mathcal{R}} z_{r, m}^{\text{c}} \geq \left(1 + \delta_{\text{c}} \right) \mu_{\text{c}} \right],\\
		& \hspace*{1.2cm}\text{where } \mu_{\text{c}} = \sum_{r \in \mathcal{R}} \frac{\tilde{x}_{r, m} c_r}{\alpha_{\text{c}}} \text{ and } z_{r, m}^{\text{c}} = \frac{\hat{x}_{r, m} c_r}{\alpha_{\text{c}}}.\notag
	\end{align}

	Now, for a given MEC $m \in \mathcal{M}$, $z_{r, m}^{\text{c}} \in [0, 1]$, $\forall r \in \mathcal{R}$, are independent random variables (RVs) with expected total value $\mathbb{E}\left[ \sum_{r \in \mathcal{R}} z_{r, m}^{\text{c}} \right] = \mu_{\text{c}}$. By following the Chernoff bound (upper tail)~\cite{mitzenmacherProbabilityComputingRandomized2005}, we get
	\begin{equation*}
		\text{Pr}\left[ \sum_{r \in \mathcal{R}} z_{r, m}^{\text{c}} \geq \left(1 + \delta_{\text{c}} \right) \mu_{\text{c}} \right] \leq \text{exp}^{\frac{-\delta_{\text{c}}^2 \mu_{\text{c}}}{2 + \delta_{\text{c}}}}, \text{ which implies to}
	\end{equation*}
	\begin{equation*}
		\text{Pr}\left[ \sum_{r \in \mathcal{R}} \frac{\hat{x}_{r, m} c_r}{\alpha_{\text{c}}} \geq \left(1 + \delta_{\text{c}} \right) \sum_{r \in \mathcal{R}} \frac{\tilde{x}_{r, m} c_r}{\alpha_{\text{c}}} \right] \leq \text{exp}^{\frac{-\delta^2_{\text{c}} \mu_{\text{c}}}{2 + \delta_{\text{c}}}},
	\end{equation*}
	which is equivalent to
	\begin{equation}
		\text{Pr}\left[ \sum_{r \in \mathcal{R}} \hat{x}_{r, m} c_r \geq \left(1 + \delta_{\text{c}} \right) \sum_{r \in \mathcal{R}} \tilde{x}_{r, m} c_r \right] \leq \text{exp}^{\frac{-\delta^2_{\text{c}} \mu_{\text{c}}}{2 + \delta_{\text{c}}}}.
		\label{eqn:chernoff_bound_cpu}
	\end{equation} 

	Next, we need to find a value of $\delta_{\text{c}}$ for which the probability value quickly goes to zero as the number of requests increases. To do that, we set
	\begin{equation}
		\text{exp}^{\frac{-\delta_{\text{c}}^2 \mu_{\text{c}}}{2 + \delta_{\text{c}}}} \leq \frac{1}{R^3},
	\end{equation}
	which means $\delta_{\text{c}}$ must satisfy the below inequality:
	\begin{equation}
		\delta_{\text{c}} \geq \frac{3}{2} \frac{\ln(R)}{\mu_{\text{c}}} + \sqrt{\frac{9}{4} \left( \frac{\ln(R)}{\mu_{\text{c}}}\right)^2 + \frac{6 \ln (R)}{\mu_{\text{c}}}}.
		\label{eqn:delta_value_condition}
	\end{equation} 
	
	To hold the above condition true, $\delta_{\text{c}}$ must be as follows:
	\begin{equation}
		\delta_{\text{c}} = \frac{3 \ln \left(R\right)}{\mu_{\text{c}}} + 3
		\label{eqn:delta_value}
	\end{equation}

	Finally, we upper bound the probability that any of the MECs CPU capacity is violated using Union bound\cite{comtetAdvancedCombinatorics1974} as:
	\begin{align}
		& \text{Pr}\left[ \bigcup\limits_{m \in \mathcal{M}} \sum_{r \in \mathcal{R}} z_{r, m}^{\text{c}} \geq (1 + \delta_{\text{c}}) \mu_{\text{c}} \right]\notag\\
		& \leq \sum_{m \in \mathcal{M}} \text{Pr}\left[ \sum_{r \in \mathcal{R}} z_{r, m}^{\text{c}} \geq (1 + \delta_{\text{c}}) \mu_{\text{c}} \right]\notag\\
		& \leq \sum_{m \in \mathcal{M}} \text{Pr}\left[ \sum_{r \in \mathcal{R}} \frac{\hat{x}_{r, m} c_r}{\alpha_{\text{c}}} \geq \left(1 + \delta_{\text{c}} \right) \sum_{r \in \mathcal{R}} \frac{\tilde{x}_{r, m} c_r}{\alpha_{\text{c}}} \right] \notag \\
		& \leq M \frac{1}{R^3} \leq \frac{1}{R^2}.
		\label{eqn:union_bound_cpu}
	\end{align}
	The last inequality in~\eqref{eqn:union_bound_cpu} holds as the number of MECs $M$ is much lesser than the number of requests $R$ in practice. Therefore, the CPU capacity of any MEC $m \in \mathcal{M}$ will not exceed by more than a factor of $(1 + \delta_{\text{c}}) = \frac{3 \ln \left(R\right)}{\mu_{\text{c}}} + 4$ with high probability.
\end{proof}

\begin{mylemma}\label{lemma:RAM_capacity}
	The RAM load on an MEC $m \in \mathcal{M}$ returned by the Algorithm~\ref{Algo1:Randomized_rounding} will not exceed its capacity by more than a factor of $(1 + \delta_{\text{d}}) = \frac{3 \ln \left(R\right)}{\mu_{\text{d}}} + 4$ with high probability, where $\mu_{\text{d}} = \sum_{r \in \mathcal{R}} \frac{\tilde{x}_{r, m} d_r}{\alpha_{\text{d}}}$, and $\alpha_{\text{d}} = \max\{d_r, \forall r \in \mathcal{R}\}$.
\end{mylemma}

\begin{mylemma}\label{lemma:uplink}
	The uplink load on an MEC $m \in \mathcal{M}$ returned by the Algorithm~\ref{Algo1:Randomized_rounding} will not exceed its capacity by more than a factor of $(1 + \delta_{\text{b}}^{\text{up}}) = \frac{3 \ln \left(R\right)}{\mu_{\text{b}}^{\text{up}}} + 4$ with high probability, where $\mu_{\text{b}}^{\text{up}} = \sum_{r \in \mathcal{R}} \frac{\tilde{x}_{r, m} b_r^{\text{up}}}{\alpha_{\text{b}}^{\text{up}}}$, and $\alpha_{\text{b}}^{\text{up}} = \max\{b_r^{\text{up}}, \forall r \in \mathcal{R}\}$.
\end{mylemma}

\begin{mylemma}\label{lemma:downlink}
	The downlink load on an MEC $m \in \mathcal{M}$ returned by the Algorithm~\ref{Algo1:Randomized_rounding} will not exceed its capacity by more than a factor of $(1 + \delta_{\text{b}}^{\text{dw}}) = \frac{3 \ln \left(R\right)}{\mu_{\text{b}}^{\text{dw}}} + 4$ with high probability, where $\mu_{\text{b}}^{\text{dw}} = \sum_{r \in \mathcal{R}} \frac{\tilde{x}_{r, m} b_r^{\text{dw}}}{\alpha_{\text{b}}^{\text{dw}}}$, and $\alpha_{\text{b}}^{\text{dw}} = \max\{b_r^{\text{dw}}, \forall r \in \mathcal{R}\}$.
\end{mylemma}

\begin{proof}
	The proofs for Lemmas~\ref{lemma:RAM_capacity}, \ref{lemma:uplink}, and \ref{lemma:downlink} follow the proof of the upper bound of the CPU load in Lemma~\ref{lemma:CPU_capacity}.
\end{proof}

Now, we study the theoretical bound on the objective value.

\begin{mylemma}
	The objective value returned by Algorithm~\ref{Algo1:Randomized_rounding} is atmost $(1 - \sqrt{\frac{4 \ln(R)}{\mu_{\text{opt}}}})$ times worse than the optimal solution of the relaxed problem, where $\mu_{\text{opt}} = \sum_{r \in \mathcal{R}} \frac{\zeta_r \tilde{y}_r}{\alpha_{\text{opt}}}$ and $\alpha_{\text{opt}} = \max\{\zeta_r, \forall r \in \mathcal{R}\}$.
\end{mylemma}

\begin{proof}
	Let $\delta_{\text{opt}} = \sqrt{\frac{4 \ln(R)}{\mu_{\text{opt}}}}$. 
	Mathematically, we are interested in finding the probability that the above lemma does not hold, i.e., the following condition is true:
	\begin{equation}
		\text{Pr}\left[ \sum_{r \in \mathcal{R}} \zeta_r \hat{y}_r  \leq (1 - \delta_{\text{opt}}) \sum_{r \in \mathcal{R}}\zeta_r \tilde{y}_r \right]
		\label{eqn:probability_bound_objective}
	\end{equation}

	We will apply the Chernoff bound (lower tail)~\cite{mitzenmacherProbabilityComputingRandomized2005} to find the bound. Before that, we normalize both sides of~\eqref{eqn:probability_bound_objective} as follows:	
	\begin{align}
		& \text{Pr}\left[ \sum_{r \in \mathcal{R}} \frac{\zeta_r \hat{y}_r}{\alpha_{\text{opt}}}  \leq (1 - \delta_{\text{opt}}) \sum_{r \in \mathcal{R}} \frac{\zeta_r \tilde{y}_r}{\alpha_{\text{opt}}} \right],\\
		& \hspace*{2cm} \text{ where } \alpha_{\text{opt}} = \max\{\zeta_r, \forall r \in \mathcal{R}\}\notag,\\
		& \Rightarrow \text{Pr}\left[ \sum_{r \in \mathcal{R}} z_{r}^{\text{opt}} \leq (1 - \delta_{\text{opt}}) \mu_{\text{opt}} \right].
	\end{align}

	The values of $z_r^{\text{opt}} \in [0, 1]$, $\forall r \in \mathcal{R}$, are independent RVs, and $\mathbb{E}\left[ \sum_{r \in \mathcal{R}} z_r^{\text{opt}} \right] = \mu_{\text{opt}}$. Now, using the Chernoff bound (lower-tail)~\cite{mitzenmacherProbabilityComputingRandomized2005}, we get
	\begin{equation*}
		\text{Pr}\left[ \sum_{r \in \mathcal{R}} z_{r}^{\text{opt}} \leq (1 - \delta_{\text{opt}}) \mu_{\text{opt}} \right] \leq \text{exp}^{\frac{-\delta_{\text{opt}}^2 \mu_{\text{opt}}}{2}}, \text{ which implies}
	\end{equation*}
	\begin{equation*}
		\text{Pr}\left[ \sum_{r \in \mathcal{R}} \frac{\zeta_r \hat{y}_r}{\alpha_{\text{opt}}}  \leq (1 - \delta_{\text{opt}}) \sum_{r \in \mathcal{R}}\frac{\zeta_r \tilde{y}_r}{\alpha_{\text{opt}}} \right] \leq \text{exp}^{\frac{-\delta_{\text{opt}}^2 \mu_{\text{opt}}}{2}},
	\end{equation*}
	which is equivalent to
	\begin{equation}
		\text{Pr}\left[ \sum_{r \in \mathcal{R}} \zeta_r \hat{y}_r  \leq (1 - \delta_{\text{opt}}) \sum_{r \in \mathcal{R}}\zeta_r \tilde{y}_r \right] \leq \text{exp}^{\frac{-\delta_{\text{opt}}^2 \mu_{\text{opt}}}{2}}.
		\label{eqn:chernoof_bound_objective}
	\end{equation}
	
	Now, we upper bound the right hand side of the inequality in~\eqref{eqn:chernoof_bound_objective} by $\frac{1}{R^2}$, and we get
	
	\begin{align}
		& \text{exp}^{\frac{-\delta_{\text{opt}}^2 \mu_{\text{opt}}}{2}} \leq \frac{1}{R^2}\notag\\
		& \Rightarrow \delta_{\text{opt}} \geq \sqrt{\frac{4 \ln(R)}{\mu_{\text{opt}}}}.
	\end{align}
	Therefore, the lowest value of $\delta_{\text{opt}}$ is $\sqrt{\frac{4 \ln(R)}{\mu_{\text{opt}}}}$ for which the above condition holds. Thus, with high probability, the objective value returned by the randomized algorithm is atmost $\left(1 - \sqrt{\frac{4 \ln(R)}{\mu_{\text{opt}}}}\right)$ times worse than the the optimal solution of the relaxed problem.
\end{proof}

\subsection{Greedy Solution}\label{Secsub:greedy_approach}
The solution returned by the randomized algorithm may not be feasible due to the violation in at least one of the capacity constraints -- CPU, RAM, uplink, and downlink, as mentioned in Section~\ref{Secsub:approximated_solution}. Therefore, we propose a feasible solution for the given approximated solution. Algorithm~\ref{Algo1:Greedy_algorithm} presents a greedy approach to obtain a feasible solution. We check the over-utilized MECs and remove requests one-by-one according to their reward\footnote{Request with the smallest reward is removed first.} values until the approximated solution is feasible (refer to \textOF{Step~\ref{step:remove_requests}}). Therefore, the total objective value obtained by the greedy approach can be lower than the approximated solution. We note that any other greedy approach can be applied to obtain a feasible solution. 
\begin{algorithm}[!ht]
	\begin{algorithmic}[1]
		\State Get the solution $(\hat{\mathbf{x}}, \hat{\mathbf{y}})$ from Algorithm~\ref{Algo1:Randomized_rounding}
		
		\For{$m \in \mathcal{M}$}
			\For{$r \in \mathcal{R}$}
				\State \textOF{Calculate} CPU utilization $C_{m}^{\text{ut}}$
				\State \textOF{Calculate} RAM utilization $D_{m}^{\text{ut}}$
				\State \textOF{Calculate} uplink utilization $B_{m}^{\text{up-ut}}$
				\State \textOF{Calculate} downlink utilization $B_{m}^{\text{dw-ut}}$
			\EndFor
			\While{$C_{m}^{\text{ut}} > C_m$ \textbf{or} $D_{m}^{\text{ut}} > D_m$ \textbf{or} \newline \hspace*{2cm} $B_{m}^{\text{up-ut}} > B_m^{\text{up}}$ \textbf{or} $B_{m}^{\text{dw-ut}} > B_m^{\text{dw}}$}
				\State $\tilde{\mathcal{R}}$ $\leftarrow$ \textOF{Get} requests with $\hat{x}_{r, m} = 1$
				
				\State $\tilde{r}$ $\leftarrow$ \textOF{Get} request from $\tilde{\mathcal{R}}$ with the lowest $\zeta_r$
				
				\State \textOF{Set} $\hat{x}_{r, m} = 0, \forall m \in \mathcal{M}$, and $\hat{y}_r = 0$\label{step:remove_requests}
				
				\State $C_m^{\text{ut}} \leftarrow C_m^{\text{ut}} - c_r$ and $D_m^{\text{ut}} \leftarrow D_m^{\text{ut}} - d_r$
				
				\State $B_m^{\text{up-ut}} \leftarrow B_m^{\text{up-ut}} - b_r^{\text{up}}$ and $B_m^{\text{dw-ut}} \leftarrow B_m^{\text{dw-ut}} - b_r^{\text{dw}}$
			
			\EndWhile
		\EndFor
		
		\State \Return \textOF{Reward} $\leftarrow$ $\sum_{r \in \mathcal{R}} \zeta_r \hat{y}_r$
		
	\end{algorithmic}
	\caption{Feasible solution: Greedy algorithm}
	\label{Algo1:Greedy_algorithm}
\end{algorithm}

\subsection{Time Complexity Analysis}\label{Secsub:Time_complexity}
We analyze the time complexity of the proposed scheme, which consists of three phases \textemdash~a) solving the relaxed (linear) problem, b) randomized rounding, and c) greedy algorithm. The time complexity of solving the relaxed (linear) problem is $O(N^3)$ and for the randomized rounding is $O(N \log N)$, where $N = |\mathcal{M}| \times |\mathcal{R}|$ is the total number of variables in the relaxed problem being rounded. The greedy approach has a time complexity of $O(|\mathcal{M}| \times |\mathcal{R}|)$. Therefore, the total time complexity is: $O(N^3) + O(N \log N) + O(|\mathcal{M}| \times |\mathcal{R}|)$.

\section{Performance Evaluation}\label{Sec:Performance_evaluation}


In this section, we evaluate the performance of the proposed scheme to show its effectiveness. Table~\ref{Table:simulation_settings} presents the parameters and their values used for the performance evaluation that are considered based on the literature~\cite{yalaLatencyAvailabilityDriven2018, pedreno-manresaNeedJointBandwidth2018, poularakisJointServicePlacement2019, zhangAdaptiveInterferenceawareVNF2019, VerticalsURLLCUse2020}.

\textbf{Network setup:} We deploy a network with 10 MECs, in which a service request can be served from any of the MECs by placing the UPFs associated with the request. We note that the UPFs associated with the two services are completely isolated and independent. The CPU and RAM capacities of an MEC are chosen at uniform random from the range specified in the Table~\ref{Table:simulation_settings}. Furthermore, the failure probabilities of MECs and UPFs are considered from~\cite{yalaLatencyAvailabilityDriven2018}, and the uplink and downlink capacities at the MECs are considered from~\cite{poularakisJointServicePlacement2019}. 

\textbf{Request generation:} We generate requests considering the UPFs required to support different use-case scenarios~\cite{pedreno-manresaNeedJointBandwidth2018}. Table~\ref{Table:CPU_RAM_VNF} presents the CPU and RAM requirements by different types of UPFs~\cite{pedreno-manresaNeedJointBandwidth2018}. Furthermore, we categorize the UPFs into two sets \textemdash~T1 and T2. For a request, all UPFs from T1 are always required to ensure network security. In contrast, we select two UPFs from T2 in a uniform random manner using \textOF{random.sample} method available in Python. Based on the selected UPFs for the request, the CPU and RAM requirements are calculated from Table~\ref{Table:CPU_RAM_VNF}. The uplink and downlink bandwidth requirements of a service request are shown in Table~\ref{Table:simulation_settings}. We note that the traffic associated with the request follows the desired service function chain (SFC) during execution. The desired SFC can be easily maintained as all the required UPFs are placed on the same machine.

\begin{table}[!ht]
	\caption{Simulation settings}
	\label{Table:simulation_settings}
	\begin{tabular}{|l|l|}
		\hline
		\textbf{Parameter}                                                             & \textbf{Value}                                        \\ \hline
		Number of MECs                                                                 & 10                                                    \\ \hline
		CPU at each MEC                                                                & {[}32, 56{]} core                                     \\ \hline
		RAM at each CPU                                                                & [32, 80] GB                                     \\ \hline
		Uplink capacity of each MEC & 75 Mbps~\cite{poularakisJointServicePlacement2019} \\ \hline
		Downlink capacity of each MEC & 250 Mbps~\cite{poularakisJointServicePlacement2019} \\ \hline
		Failure of a VNF ($\epsilon_{\text{v}}$) \cite{yalaLatencyAvailabilityDriven2018}              & 0.001                                                 \\ \hline
		Failure of a PM ($\epsilon_{\text{p}}$) \cite{yalaLatencyAvailabilityDriven2018} & 0.004                                                 \\ \hline
		Number of requests                                                             & \{30, 35, 40, 50, 60\}                                   \\ \hline
		Avail. requirements ($1 - \epsilon_r$)~\cite{VerticalsURLLCUse2020} & \{0.99, 0.999, 0.9999\} \\ \hline
		Reward ($\zeta_r$) & [6, 8] $\times$ $(1 - \epsilon_r)$ \\ \hline
		CPU and RAM requirements~\cite{pedreno-manresaNeedJointBandwidth2018}            & Table~\ref{Table:CPU_RAM_VNF} \\ \hline
		Uplink requirements & [6, 15] Mbps \\ \hline
		Downlink requirements & [20, 40] Mbps\\ \hline
	\end{tabular}
\end{table}

\begin{table}[!ht]
	\centering
	\caption{UPFs requirements \cite{pedreno-manresaNeedJointBandwidth2018}}
	\label{Table:CPU_RAM_VNF}
	\begin{tabular}{lcccc}
		\hline
		\multicolumn{1}{|c|}{\textbf{VNF}$^{1}$} & \multicolumn{1}{c|}{\textbf{CPU (Core)}} & \multicolumn{1}{c|}{\textbf{RAM (GB)}} & \multicolumn{1}{c|}{\textbf{T1}}                                                          & \multicolumn{1}{c|}{\textbf{T2}}                                                                    \\ \hline
		\multicolumn{1}{|l|}{IDPS}         & \multicolumn{1}{c|}{2}                   & \multicolumn{1}{c|}{2}                 & \multicolumn{1}{c|}{\multirow{6}{*}{\begin{tabular}[c]{@{}c@{}}NAT\\ \\ FW\end{tabular}}} & \multicolumn{1}{c|}{\multirow{6}{*}{\begin{tabular}[c]{@{}c@{}}IDPS\\ TM\\ VOC\\ WOC\end{tabular}}} \\ \cline{1-3}
		\multicolumn{1}{|l|}{FW}           & \multicolumn{1}{c|}{2}                   & \multicolumn{1}{c|}{3}                 & \multicolumn{1}{c|}{}                                                                     & \multicolumn{1}{c|}{}                                                                               \\ \cline{1-3}
		\multicolumn{1}{|l|}{NAT}          & \multicolumn{1}{c|}{1}                   & \multicolumn{1}{c|}{1}                 & \multicolumn{1}{c|}{}                                                                     & \multicolumn{1}{c|}{}                                                                               \\ \cline{1-3}
		\multicolumn{1}{|l|}{TM}           & \multicolumn{1}{c|}{1}                   & \multicolumn{1}{c|}{3}                 & \multicolumn{1}{c|}{}                                                                     & \multicolumn{1}{c|}{}                                                                               \\ \cline{1-3}
		\multicolumn{1}{|l|}{VOC}          & \multicolumn{1}{c|}{2}                   & \multicolumn{1}{c|}{2}                 & \multicolumn{1}{c|}{}                                                                     & \multicolumn{1}{c|}{}                                                                               \\ \cline{1-3}
		\multicolumn{1}{|l|}{WOC}          & \multicolumn{1}{c|}{1}                   & \multicolumn{1}{c|}{2}                 & \multicolumn{1}{c|}{}                                                                     & \multicolumn{1}{c|}{}                                                                               \\ \hline
		\multicolumn{5}{l}{\begin{tabular}[c]{@{}l@{}}$^1$IDPS: Intrusion detection and prevention system\\ FW: Firewall; NAT: Network address translation\\ TM: Traffic monitor; VOC: Video optimizer controller\\ WOC: WAN optimizer controller\end{tabular}}                                                                     
	\end{tabular}
\end{table}

\subsection{Results and Discussion}\label{Secsub:Results_discussion}
We present two variations of the proposed scheme \textemdash~randomized rounding (Algorithm~\ref{Algo1:Randomized_rounding}) and greedy algorithm (Algorithm~\ref{Algo1:Greedy_algorithm}). We compare the proposed scheme with the optimal solution to the relaxed problem. We also compare the proposed scheme with the schemes that do not consider the availability requirement, such as~\cite{ yalaLatencyAvailabilityDriven2018, poularakisJointServicePlacement2019, poularakisServicePlacementRequest2020, yangDelayawareVirtualNetwork2021}. We note that the authors in~\cite{ yalaLatencyAvailabilityDriven2018} considered the availability requirement by placing the VNFs at the central cloud, which may not be suitable for uRLLC applications, considering their stringent latency requirements. For the existing schemes that do not consider the availability requirements, we remove the constraint~\eqref{eqn:constraint_redundancy} to eliminate the availability factor, meaning that each request is assigned without considering redundancy. As a result, a request cannot be served in case of failures in the associated VNFs. Henceforth, we refer \textOF{LR}, \textOF{RR}, \textOF{Greedy}, and \textOF{Wo-Avl} to present the optimal solution, the randomized rounding approach, the proposed greedy approach, and the schemes that do not consider the availability, respectively.

We take the average of 50 runs of the experiment and present the results with 95\% confidence interval~\cite{hackshawStatisticalFormulaeCalculating2009}. We consider three performance metrics \textemdash~total reward, resource utilization, and percentage of requests served. The total reward is presented with varying numbers of requests and networking resources. The resource utilization includes the percentage of CPU, RAM, uplink, and downlink capacity utilization at the MECs.

\subsubsection{Total Reward}\label{Secsubsub:result_reward}
\begin{figure}[!ht]
	\centering
	\includegraphics[scale=0.55]{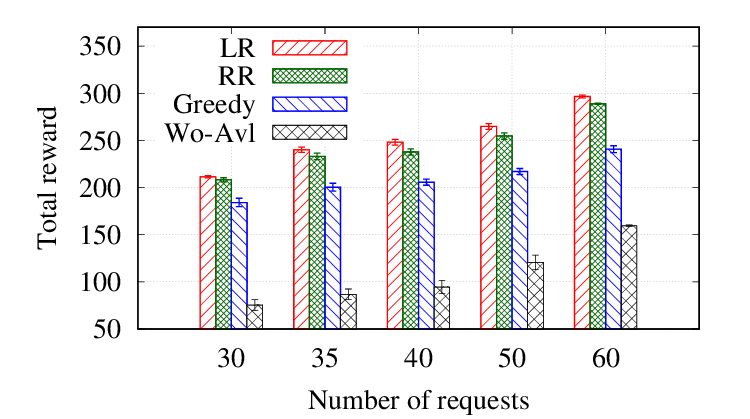}
	\caption{Reward with number of requests}
	\label{fig:result_reward_requests}
\end{figure}

The service provider's objective is to maximize the total reward by serving the requests, as mentioned in Section~\ref{Sec:System_model}. Figure~\ref{fig:result_reward_requests} presents the total reward obtained by the service provider with varying the numbers of requests using \textOF{LR}, \textOF{RR}, \textOF{Greedy}, and \textOF{Wo-Avl}. It is evident that \textOF{RR} yields a competitive reward when compared to \textOF{LR}. Furthermore, \textOF{Greedy} also yields a competitive performance to both \textOF{LR} and \textOF{RR}. In particular, the total rewards obtained by \textOF{RR} and \textOF{Greedy} are within $7\%$ and $17\%$, respectively, when compared to the optimal solution of the relaxed problem (\textOF{LR}). We note that \textOF{Greedy} achieves a lower reward than \textOF{RR}, as some of the requests are dropped to make the solution feasible, as presented in Section~\ref{Secsub:greedy_approach}. Whereas \textOF{Wo-Avl} achieves 62\%, 60\%, and 51\% lower rewards than \textOF{LR}, \textOF{RR}, and \textOF{Greedy}, respectively. This is because the lack of redundancy leads to a higher probability of service failures, thereby reducing the overall number of successfully served requests. 

\begin{figure}[!ht]
	\centering
	\subfigure[CPU]{
		\includegraphics[scale=0.38]{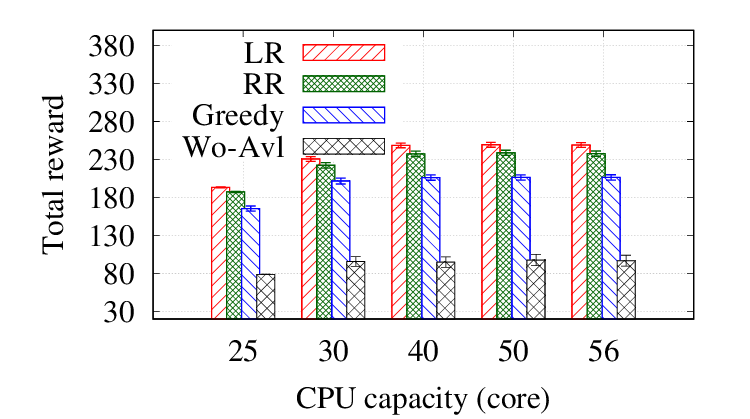}
		\label{fig:result_reward_cpu}	
	}\hspace*{-2em}
	\subfigure[RAM]{
		\includegraphics[scale=0.38]{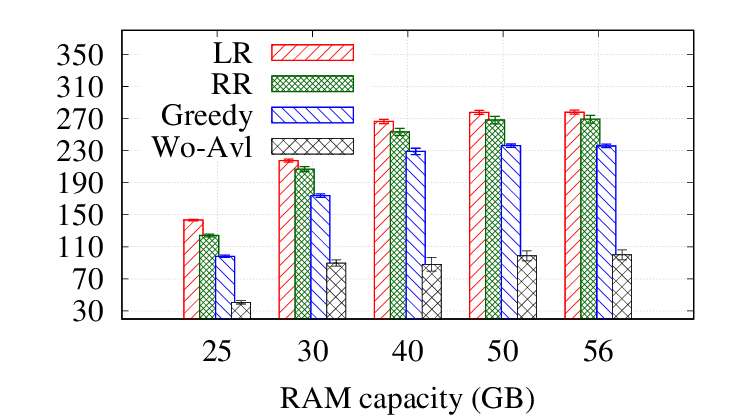}
		\label{fig:result_reward_ram}	
	}
	\subfigure[Uplink]{
		\includegraphics[scale=0.38]{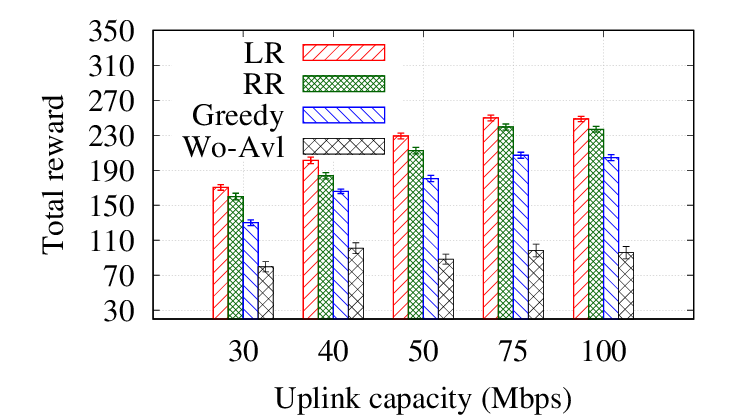}
		\label{fig:result_reward_uplink}	
	}\hspace*{-2em}
	\subfigure[Downlink]{
		\includegraphics[scale=0.38]{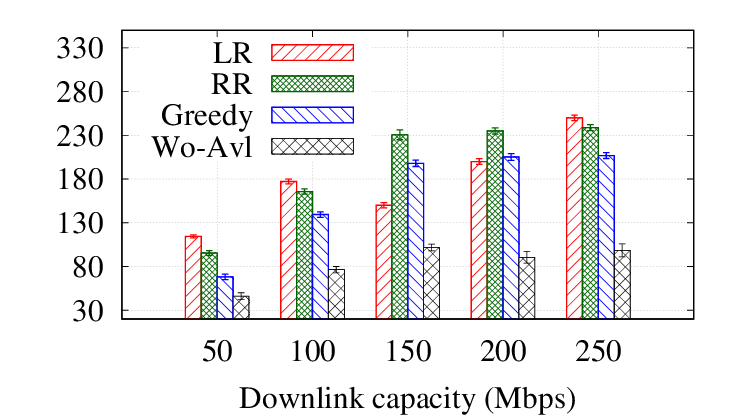}
		\label{fig:result_reward_downlink}	
	}
	\caption{Impact of network resources on total reward ($C_m = 40$, $D_m = 48$, $B_m^{\text{up}} = 75$, and $B_m^{\text{dw}} = 250$, $R = 50$)}
	\label{fig:result_impact_reward_resources}
\end{figure}

To understand the impact of different networking resources at the MECs, we present the total reward with varying networking resources in Figure~\ref{fig:result_impact_reward_resources} using \textOF{LR}, \textOF{RR}, \textOF{Greedy}, and \textOF{Wo-Avl}. Figures~\ref{fig:result_reward_cpu}, \ref{fig:result_reward_ram}, \ref{fig:result_reward_uplink}, and \ref{fig:result_reward_downlink} present the total reward with varying CPU, RAM, uplink, and downlink resources, respectively, one at a time, while the other resources are fixed. In other words, the RAM, uplink, and downlink resources are fixed when CPU capacity is varied. The specific values considered for this evaluation is mentioned in Figure~\ref{fig:result_impact_reward_resources}. Similar to Figure~\ref{fig:result_reward_requests}, we see that \textOF{RR} and \textOF{Greedy} yield competitive performance to \textOF{LR}. Whereas \textOF{Wo-Avl} yields the lowest reward compared to the proposed approaches. Furthermore, it is observed that the total reward does not increase even after increasing the resources after a certain point. This is because more requests cannot be served due to the limitation in the other networking resources.

%
%

\subsubsection{Requests Served}\label{Secsubsub:result_requests_served}
\begin{figure}[!ht]
	\centering
	\includegraphics[scale=0.55]{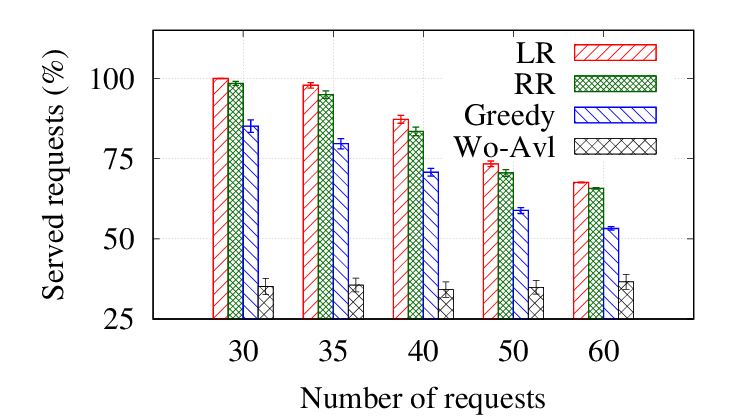}
	\caption{Percentage of served requests with number of requests}
	\label{fig:result_served_requests}
\end{figure}

Considering diverse applications, a service provider can also be interested in serving as many requests as possible. Considering this, we evaluate the percentage of requests served at the MECs, as shown in Figure~\ref{fig:result_served_requests}. We see that the proposed schemes, \textOF{RR} and \textOF{Greedy}, yield competitive performance to the optimal solution \textOF{LR}. In particular, the percentages of requests served by \textOF{RR} and \textOF{Greedy} are within $6\%$ and $18\%$ to \textOF{LR}. However, the percentages of served requests using \textOF{Wo-Avl} is reduced by $57\%$ and $49\%$ in comparison to \textOF{LR} and \textOF{Greedy}, respectively. Therefore, it is evident that the consideration of the availability requirement of uRLLC requests plays an important role in serving the requests. Furthermore, it is observed that the percentage of served requests decreases with an increase in the number of requests. This is because of the limited networking resources available at the MECs, which, in turn, reduces the ratio between served and total requests.

\subsubsection{Resource Utilization}\label{Secsubsub:result_utilization}

\begin{figure}[!ht]
	\centering
	\includegraphics[scale=0.17]{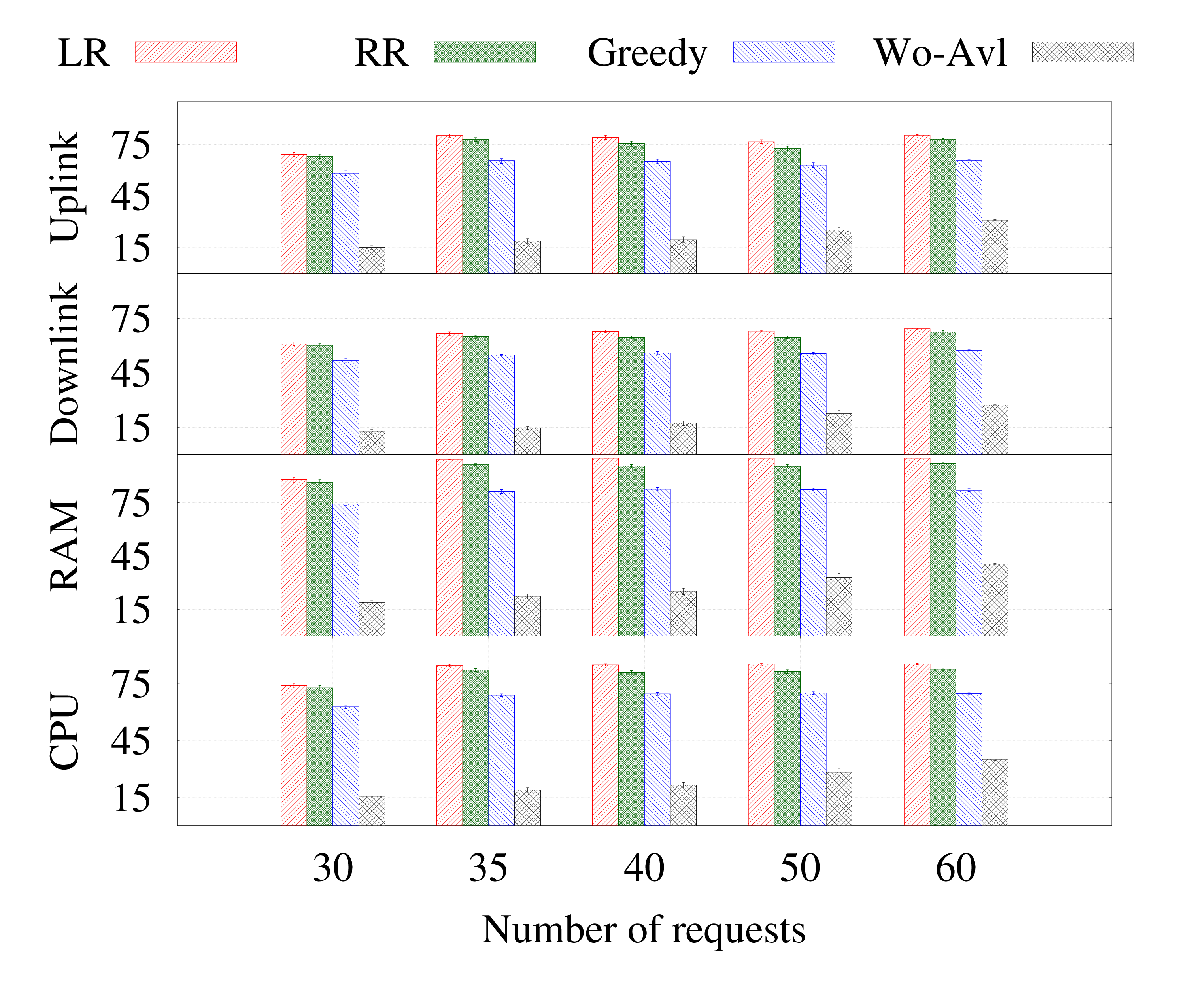}
	\caption{Percentage of resource utilization with number of requests}
	\label{fig:result_resource_utilization}
\end{figure}

As mentioned in Section~\ref{Sec:Introduction}, the resources available at MECs are limited compared to a central cloud. Therefore, we measure the resource utilization at the MECs in terms of CPU, RAM, uplink, and downlink capacity utilization. Figure~\ref{fig:result_resource_utilization} shows the percentages of resource utilizations at the MECs with varying numbers of requests. In all cases, we see that resource utilization increases with an increase in the number of requests for all schemes. This is because more requests are served by the MECs to maximize the reward. However, the utilization at the MECs gets saturated after a certain number of requests. Furthermore, we note that resource utilization for \textOF{RR} and \textOF{Greedy} is lower by $8\%$ and $18\%$, respectively, than that of the \textOF{LR}, while yielding competitive rewards to the latter. This is because some requests are not served in the rounding procedure, further in the \textOF{Greedy} approach. Whereas none of the requests with redundant VNF requirements are served using \textOF{Wo-Avl} (refer to Figure~\ref{fig:result_served_requests}), which reduces the percentage of resource utilization.

\subsubsection{Computation Time}\label{Secsubsub:result_computation_time}
We compare the proposed schemes with the integer solution obtained by solving the original optimization problem~\eqref{eqn:objective_optimization} in terms of computation time using \textOF{IBM CPLEX}~\cite{CPLEXCPOptimizer2020}. As the problem is NP-hard, the computation time for the integer solution, called \textOF{Integer}, is higher by two orders of magnitude than the other schemes -- \textOF{RR}, \textOF{Greedy}, and \textOF{Wo-Avl}. Furthermore, the computation time of \textOF{Greedy} is slightly higher than \textOF{RR}. This is due to the additional computation required by \textOF{Greedy} after obtaining the rounding solution \textOF{RR}. The computation time of \textOF{Wo-Avl} is slightly lower than \textOF{RR} and \textOF{Greedy}. This is due to reduced computational complexity by ignoring the availability requirements.
\begin{figure}[!ht]
	\centering
	\includegraphics[scale=0.55]{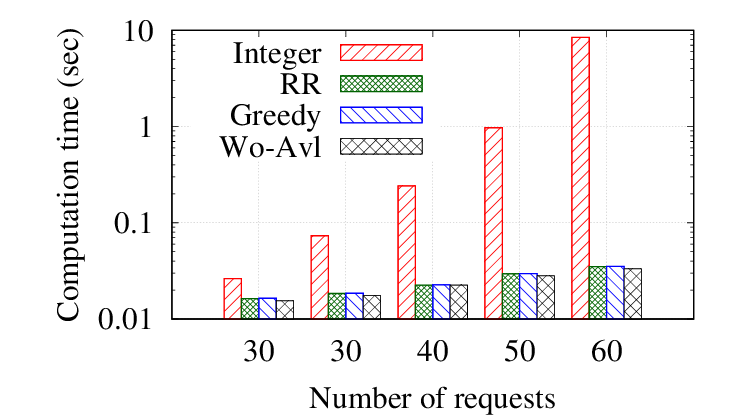}
	\caption{Computation time}
	\label{fig:result_computation_time}
\end{figure}

In summary, the proposed approaches, \textOF{RR} and \textOF{Greedy}, yield competitive performance to the optimal solution to the relaxed optimization problem, \textOF{LR}, while requiring polynomial computation time when compared to the \textOF{Integer} solution. Furthermore, the proposed \textOF{Greedy} approach outperforms the existing schemes, \textOF{Wo-Avl}, that do not consider the availability requirements of uRLLC requests.

\section{Software Prototype Implementation: Redundant UPF Placement}\label{Sec:Prototype_Implementation}
\begin{figure}[!ht]
	\centering
	\includegraphics[scale=1.3]{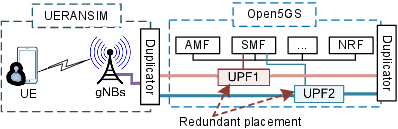}
	\caption{Prototype implementation of redundant UPF placement in 5G network}
	\label{fig:prototype}
\end{figure}

We implemented a prototype of the entire 5G system using Open5GS (\url{https://open5gs.org/}) and UERANSIM (\url{https://github.com/aligungr/UERANSIM}). The Open5GS and UERANSIM are used to deploy the 5G core network and the radio access network (RAN), respectively. We use the \textOF{Linux Ubuntu 20.04 LTS} platform to deploy the 5G network using Open5GS and UERANSIM. Figure~\ref{fig:prototype} shows the overview of the implementation, where UPF1 and UPF2 are redundantly placed in the network. All incoming and outgoing traffic is duplicated and forwarded through both UPFs.

\begin{figure}[!ht]
	\centering
	\subfigure[Packet delivery ratio]{
		\includegraphics[scale=0.5]{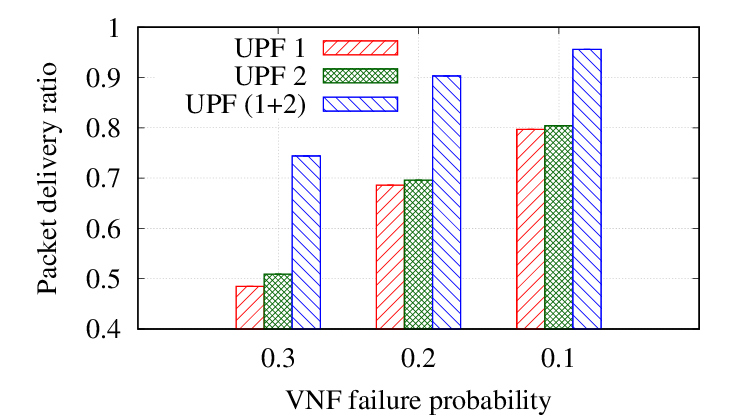}
		\label{fig:result_prototype_pdr}
	}
	\subfigure[Latency]{
		\includegraphics[scale=0.5]{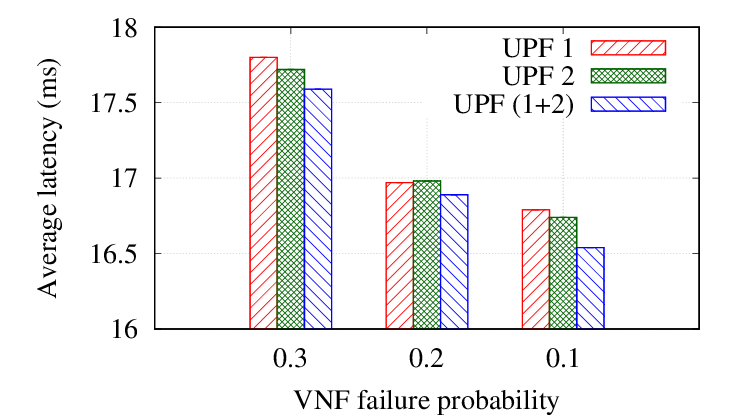}
		\label{fig:result_prototype_latency}
	}
	\caption{Network performance with different failure probabilities of UPFs}
	\label{fig:result_prototype}
\end{figure}

To evaluate the performance on packet delivery ratio (PDR) with different failure probabilities of the UPFs, we enable and disable the packet forwarding rule available in Linux. Figure~\ref{fig:result_prototype_pdr} shows the packet delivery ratio with and without redundant placement. It is evident that PDR increases with redundant placement compared to the non-redundant placement. Furthermore, we also measure the average latency in the presence of redundant UPFs. Figure~\ref{fig:result_prototype_latency} shows the average latency experienced by network traffic using single and redundant UPFs with different failure probabilities. Similar to the PDR, it is evident that latency can also be minimized with redundant placement of UPFs in the network. However, the PDR and latency are improved by incurring additional resources at the MECs in the network. Therefore, the redundant UPFs can be used based on the availability and latency requirements of the underlying applications.

\section{Conclusion}\label{Sec:Conclusion}
In this paper, we studied the VNF placement problem in MEC-enabled 5G networks while considering the reliability and latency requirements of uRLLC applications. We proposed an approximation algorithm based on the randomized rounding techniques to solve the NP-hard optimization problem in polynomial time. Furthermore, we proved the theoretical bounds of the proposed solution with respect to the optimal solution to the relaxed optimization problem. Finally, we presented simulation results to show the efficacy of the proposed approach. We also presented experiment results on packet delivery ratio and latency by implementing a software prototype of the 5G network using open-source software tools.

In this work, we considered that UPFs are independent and isolated from one service-type to another. However, a UPF can be shared among multiple services/applications to minimize CAPEX and OPEX. As a result, it may lead to compromised QoS for the underlying applications. As a future research direction, we are interested in studying the impact on the performance when UPFs are shared among multiple service-types and proposing QoS-aware service provisioning in the presence of shared UPFs.

\bibliographystyle{IEEEtran}
\bibliography{VNF_placement}

\end{document}